\documentclass[aps,superscriptaddress,floatfix]{revtex4}%
\usepackage{amsfonts}
\usepackage{amsmath}
\usepackage{amssymb}
\usepackage{graphicx}
\usepackage{graphicx}%
\begin{document}
\title{Application of the low finesse frequency comb for high resolution spectroscopy}
\author{R. N. Shakhmuratov$^{\star}$}
\affiliation{Kazan Physical-Technical Institute, Russian Academy of Sciences,
10/7 Sibirsky Trakt, Kazan 420029 Russia}
\affiliation{Kazan Federal University, 18 Kremlyovskaya Street, Kazan 420008 Russia}
\author{F. G. Vagizov}
\affiliation{Kazan Federal University, 18 Kremlyovskaya Street, Kazan 420008 Russia}
\affiliation{Institute for Quantum Studies and Engineering and Department of Physics and
Astronomy, TAMU, College Station, Texas 77843-4242, USA}
\author{Marlan O. Scully}
\affiliation{Institute for Quantum Studies and Engineering and Department of Physics and
Astronomy, TAMU, College Station, Texas 77843-4242, USA}
\author{Olga Kocharovskaya}
\affiliation{Institute for Quantum Studies and Engineering and Department of Physics and
Astronomy, TAMU, College Station, Texas 77843-4242, USA}
\date{{ \today}}

\begin{abstract}
High finesse frequency combs (HFC) with large ratio of the frequency spacing
to the width of the spectral components have demonstrated remarkable results
in many applications such as precision spectroscopy and metrology. We found
that low finesse frequency combs having very small ratio of the frequency
spacing to the width of the spectral components are more sensitive to the
exact resonance with absorber than HFC. Our method is based on time domain
measurements reviling oscillations of the radiation intensity after passing
through an optically thick absorber. Fourier analysis of the oscillations
allows to reconstruct the spectral content of the comb. If the central
component of the incident comb is in exact resonance with the single line
absorber, the contribution of the first sideband frequency into oscillations
is exactly zero. We demonstrate this technique with gamma-photon absorbtion by
M\"{o}ssbauer nuclei providing the spectral resolution beyond the natural broadening.

\end{abstract}
\maketitle

Techniques using femtosecond-laser frequency combs allow to measure extremely
narrow optical resonances with high resolution \cite{Hansch2002,Cundiff2003}.
This is achieved by precision laser stabilization and mode-locked lasers
generating the pulse train with broad frequency spectrum consisting of a
discrete, regularly spaced series of sharp lines, known as a frequency comb.
The absolute frequency of all the comb lines can be determined with high
precision. The repetition rate of the femtosecond-laser pulses is usually
ranged between 50-1000 MHz, which makes accessible practically all the optical
frequencies covered by the laser spectrum. Phase stabilization of the pulses
is capable to make a relative linewidth of the comb components to be subhertz
to sub-mHz \cite{Diddams2004,Ye2008} and absolute spectral width of the lines
in the comb as narrow as 1 Hz or even smaller over an octave spectrum
\cite{Ma2013}. Optical frequency combs have wide applications including
precision spectroscopical measurements
\cite{Hansch1999,Hansch2000,Cundiff2000,Hansch2000_I,Udem2001}, all-optical
atomic clocks \cite{Wineland2001,Hall2001,Hansch2002}, measurement of the
atomic transition linewidth and population transfer by the transient coherent
accumulation effect \cite{Ye2006}, and astronomical spectrographs calibration
for the observation of extremely small Doppler velocity drifts \cite{Udem2008}.

Broadband high-resolution X-ray frequency combs were proposed to generate by
the X-ray pulse shaping method, which imprints a comb on the excited
transition with a high photon energy by the optical-frequency comb laser
driving the transition between the metastable and excited states
\cite{Keitel2014,Pfeifer2014}. Enabling this technique in the X-ray domain is
expected to result in wide-range applications, such as more precise tests of
astrophysical models, quantum electrodynamics, and the variability of
fundamental constants. \begin{figure}[ptb]
\resizebox{0.45\textwidth}{!}{\includegraphics{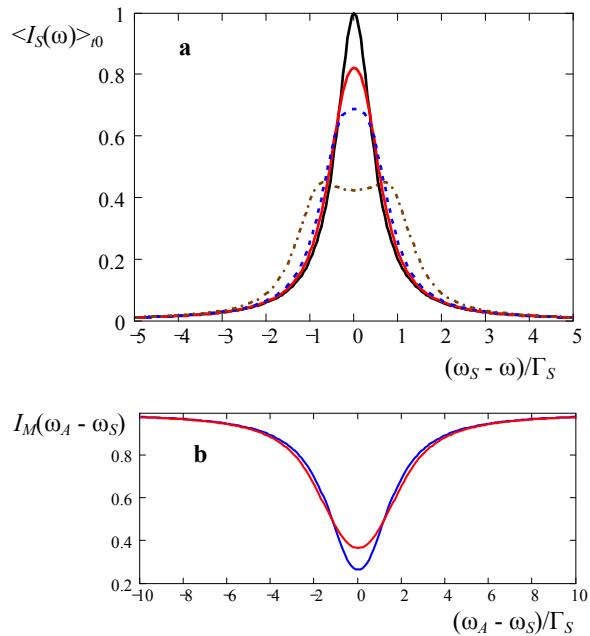}}\caption{
\textbf{Power spectrum of the source radiation field, seen by the vibrated
absorber, and absorption spectrum after passing through the absorber.}
\textbf{a,} Power spectrum, normalized to the peak intensity. \textbf{b,} The
absorption spectrum, normalized to unity. The vibration frequency is
$\Omega=\Gamma_{S}/5$. The modulation index $a$ is $0$ (solid black line), $2$
(solid red line), $3$ (blue dashed line), and 6 (brown dash-dotted line) in
\textbf{a}. The modulation index $a$ is $0$ (blue line) and 6 (red line) in
\textbf{b}. The optical thickness of the absorber is $T_{a}=\alpha(0)l=5.2$.
The linewidth of the absorber $\Gamma_{A}$ and spectral components of the
source $\Gamma_{S}$ are equal to $\Gamma_{0}$, which is defined by the lifetime
of the excited state nucleus (141 ns).}%
\label{fig:1}%
\end{figure}

Gamma-ray frequency combs were generated much earlier by Doppler modulation of
the radiation frequency, induced by mechanical vibration of the source or
resonant absorber
\cite{Cranshaw,Ruby,Kornfeld,Mishroy,Walker,Mkrtchyan77,Perlow,Monahan,
Mkrtchyan79,Tsankov,Shvydko89,Shvydko92}. They were observed in frequency
domain and appear only if the source and absorber were used in couple.
Contrary to the optical and X-ray combs, discussed above, gamma-ray frequency
combs do not produce sharp, short pulses in time domain, except the cases if
the specific conditions are satisfied \cite{Vagizov,Shakhmuratov15}.

Gamma-ray frequency combs with high finesse $F\gg1$, where $F$ is the ratio of
the comb-tooth spacing to the tooth width, demonstrated that in many cases
determination of small energy shifts between the source and absorber can be
made more accurately in time domain by transient and high-frequency modulation
techniques than by conventional methods in frequency domain
\cite{Perlow,Monahan,Helisto1984}. We have to emphasize that in gamma domain
even standard spectroscopic measurements with such a popular M\"{o}ssbauer
isotopes as $^{57}$Fe and $^{67}$Zn have already demonstrated extremely high
frequency resolution in measurements of gravitational red-shift
\cite{Pound1965,Potzel1992}. This is because the quality factor $Q$, which is
the ratio of the resonance frequency to the linewidth, is very high for these
nuclei. For example, 14.4 keV transition in $^{57}$Fe has $Q=3\times10^{12}$
and 93.3 keV resonance in $^{67}$Zn demonstrates $Q=1.8\times10^{15}$.
Appropriate sources emitting resonant or very close to resonance $\gamma
$-photons with high $Q$ are available for both nuclei. They are $^{57}$Co for
$^{57}$Fe and $^{67}$Ga for $^{67}$Zn.

Here we show that a low finesse comb (LFC) with $F\ll1$ is more sensitive to
the small resonant detuning between the fundamental of the radiation field and
the absorber compared with high finesse comb (HFC).

The basic idea of the modulation technique in gamma-domain is that the
vibration of an absorber leads to a periodic modulation of the resonant
nuclear transition frequency with respect to the frequency of the incident
photons owing to the Doppler effect. This modulation induces coherent Raman
scattering of the incident radiation in the forward direction
transforming\ quasi-monochromatic field into a frequency comb at the exit of
the absorber \cite{Shvydko92}. The relative amplitudes and phases of the
produced spectral components are defined by the vibration amplitude $d$ and
frequency $\Omega$, the detuning of the central frequency of the radiation
source $\omega_{S}$ from the resonant frequency of the absorber $\omega_{A}$,
the linewidths of the source $\Gamma_{S}$ and absorber $\Gamma_{A}$, and the
absorber optical depth $T_{A}$. To describe the transformation of the
quasi-monochromatic radiation field into a frequency comb it is convenient to
consider the interaction of the field with nuclei in the reference frame
rigidly bounded to the piston-like vibrated absorber. There, nuclei `see' the
quasi-monochromatic source radiation with main frequency $\omega_{S}$ as
polychromatic radiation with a set of spectral lines $\omega_{S}\pm n\Omega$
($n=0,\pm1,\pm2,...$) spaced apart at distances that are multiples of the
oscillation frequency. The intensity of the $n$th sideband is given by the
square of the Bessel function $J_{n}^{2}(a)$, here $a=2\pi d/\lambda$ is the
modulation index of the field phase $\varphi(t)=a\sin(\Omega t)$, $d$ is the
amplitude of the vibration, and $\lambda$ is the wavelength of the radiation.

If the modulation frequency $\Omega$ is much lager than $\Gamma_{S}$, the
power spectrum of the radiation field, seen by the absorber nuclei,
demonstrates HFC ($F=\Omega/\Gamma_{S}\gg1$). It is observed in many
M\"{o}ssbauer experiments
\cite{Cranshaw,Ruby,Kornfeld,Mishroy,Walker,Mkrtchyan77,Perlow,Monahan,
Mkrtchyan79,Tsankov,Shvydko89,Shvydko92} by transmitting the radiation field
through a single line absorber with resonant frequency $\omega_{A}$. The
carrier frequency of the radiation frequency of the source $\omega_{S}$ is
changed by a constant velocity Doppler shift. The intensity of the transmitted
radiation, showing a frequency-comb M\"{o}ssbauer spectrum, is described by
equation%
\begin{equation}
I_{M}(\omega_{A}-\omega_{S})=%
{\displaystyle\int\limits_{-\infty}^{+\infty}}
\left\langle I_{S}(\omega)\right\rangle _{t_{0}}e^{-\alpha(\omega_{A}%
-\omega)l}d\omega,\label{Eq1}%
\end{equation}
where $\alpha(\omega_{A}-\omega)$ is the frequency dependent absorption
coefficient of the single line absorber, $l$ is the absorber thickness, and
$\left\langle I_{S}(\omega)\right\rangle _{t_{0}}$ is the power spectrum of
the radiation field seen by the vibrated nuclei. Here, the power spectrum is
averaged over the random time of photon emission $t_{0}$ (see Supplementary
information for details).

Frequency-domain M\"{o}ssbauer spectrum is measured by counting the number of
photons, detected within long time windows of the same duration for all
resonant detunings, which are varied by changing the value of a constant
velocity of the M\"{o}ssbauer drive moving the source. Time windows are not
synchronized with the mechanical vibration and their duration $T_{\text{w}}$
is much longer than the oscillation period $T_{\text{osc}}=2\pi/\Omega$.

If $F\ll1$, the spectral components of the frequency comb, seen by the
absorber nuclei, overlap resulting in the spectrum broadening of the radiation
field (see Fig. 1a). Therefore M\"{o}ssbauer spectra for LFC show only the
line broadening with increase of the modulation index $a$, see Fig. 1b.

If time windows of the photon-count collection are synchronized with the phase
oscillations and duration of the time-windows $T_{\text{w}}$ is much shorter
than the oscillation period $T_{\text{osc}}$, then one can observe time
dependence of the transmitted radiation. For HFC the number of counts $N(t)$,
proportional to the radiation intensity $I(t)$, is described by the equation
\cite{Perlow,Monahan,Helisto1984,Vagizov}%
\begin{equation}
N(t)=N_{0}%
{\displaystyle\sum_{n=0}^{\infty}}
D_{n}\cos\left[  n\Omega(t-t_{n})\right]  , \label{Eq2}%
\end{equation}
where $N_{0}$ is the number of counts without absorber, $D_{n}$ and $n\Omega
t_{n}$ are the amplitude and phase of the $n$th harmonic. Here, nonresonant
absorbtion is disregarded. Recoil processes in nuclear absorption and emission
are not taken into account assuming that recoilless fraction (Debye-Waller
factor) is $f=1$. These processes can be easily taken into account in
experimental data analysis.

If the fundamental frequency $\omega_{S}$ of the comb coincides with the
resonant frequency of the single line absorber ($\omega_{S}=\omega_{A}$), then
the amplitudes of the odd harmonics are zero, $D_{2m+1}=0$, where $m$ is
integer. They become nonzero for nonresonant excitation. For high finesse
combs the ratio of the amplitudes of the first and second harmonics
$D_{1}/D_{2}$ is linearly proportional to the resonant detuning $\Delta
=\omega_{A}-\omega_{S}$ if the value of the modulation index $a$ is not large
and the resonant detuning does not exceed the linewidth $(\Gamma_{A}%
+\Gamma_{S})/2$ \cite{Perlow,Monahan}. This dependence helps to measure the
value of small resonant detuning with high accuracy \cite{Helisto1984}. For
HFC the optimal value of the modulation index providing the best signal to
noise ratio is $a=1.08$ when the amplitude of the first harmonic $D_{1}$ takes
maximum. This is because for HFC $D_{1}$ is proportional to the product of the
amplitudes of zero and first components of the comb, i.e., to $J_{0}%
(a)J_{1}(a)$.

If one of the sidebands of the comb ($\omega_{S}\pm n\Omega$) is in resonance
with the absorber ($\omega_{S}\pm n\Omega=\omega_{A}$, $n\neq0$), then time
dependence of the radiation field shows large amplitude pulses of short
duration \cite{Vagizov,Shakhmuratov15}. High sensitivity of HFC to resonance
of its central frequency with a single line absorber and formation of short
pulses if the sidebands are in resonance are explained by the interference of
the spectral components of the comb, which are changed after passing through
the absorber. In this paper we show that LFC is much more sensitive to the
resonance of the central component of the comb with the single line absorber.
This sensitivity can be explained by the following arguments.

Since the radiation intensity is $I(t)=E(t)E^{\ast}(t)$, which is the product
of the complex conjugated amplitudes $E(t)$ containing the exponential phase
factor $\exp[i\varphi(t)]$, the time dependent phase $\varphi(t)$ of the field
amplitude does not lead to the additional time dependence of the intensity.
This fact, resulting from the simple relation $e^{i\varphi(t)}e^{-i\varphi(t)}=1$,
can be explained by a particular interference of the spectral components of
this specific frequency comb. For example, only zero frequency spectral
component is present in the intensity, $I(t)=\mathtt{I}_{0}$, since all zero
spectral components, resulting from the products $e^{in\Omega t}e^{-in\Omega t}$
with $n=0,\pm1,\pm2...$, are summed up with the equal weights as%
\begin{equation}%
{\displaystyle\sum\limits_{n=-\infty}^{+\infty}}
J_{n}^{2}(a)=1, \label{Eq3}%
\end{equation}
while the first harmonic $I_{1}(t)e^{i\Omega t}$ is zero because this sum
gives%
\begin{equation}%
{\displaystyle\sum\limits_{n=-\infty}^{+\infty}}
J_{n}(a)J_{n-1}(a)=0. \label{Eq4}%
\end{equation}
The same is true for all higher frequency components. This fragile balance
between spectral components is easily broken after passing through the single
line absorber. We found that for LFC the absorber of thickness $l$ satisfying
the relation $\alpha(0)l=1$ is already capable to produce noticeable time
oscillations of the intensity. These oscillations are also described by Eq.
(\ref{Eq2}) whose coefficients can be calculated by the method developed in
\cite{Monahan,Vagizov} (see details in Supplementary information). In contrast
to HFC ($\Omega\gg\Gamma_{S}$), LFC ($\Omega\ll\Gamma_{S}$) becomes sensitive
to exact resonance if effective halfwidth of the comb $a\Omega$ is nearly
equal to the width of the absorption line $\Gamma_{A}$. Since for LFS $a\gg1$,
much more spectral components [$J_{n}(a)J_{n+1}(a)$ with $n=0,\pm 1,...,\pm a$]
participate in the interference compared with HFC. Thus, the spectral content
of the intensity oscillations becomes more sensitive to the resonant detuning.
\begin{figure}[ptb]
\resizebox{0.45\textwidth}{!}{\includegraphics{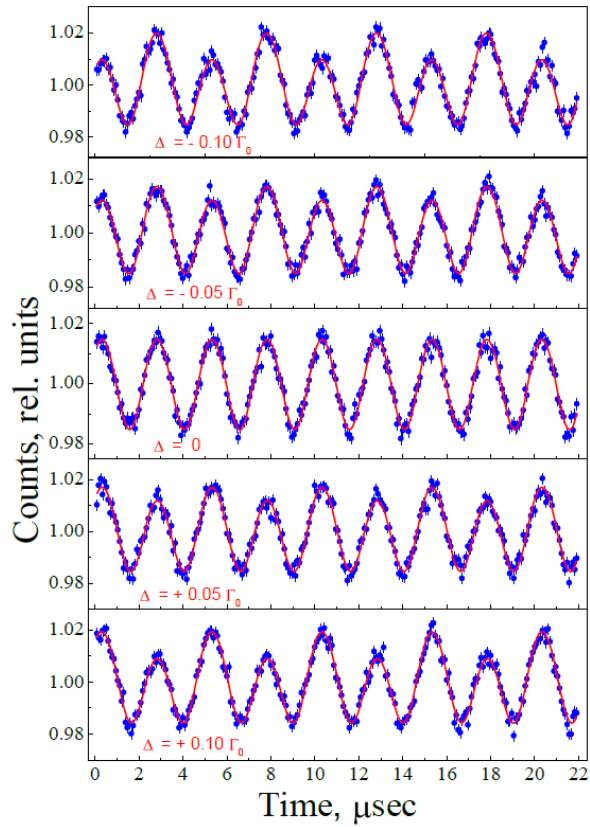}}\caption{\textbf{Time
dependence of the photon counts $N(t)$ for different values of the resonant
detuning.} The number of counts (in relative units) are normalized to the mean
value at exact resonance. The value of the detuning in units of $\Gamma_{0}$ is
indicated in each panel. The dots are experimental points and solid line is
a theoretical fitting.}%
\label{fig:2}%
\end{figure}
\begin{figure}[ptb]
\resizebox{0.4\textwidth}{!}{\includegraphics{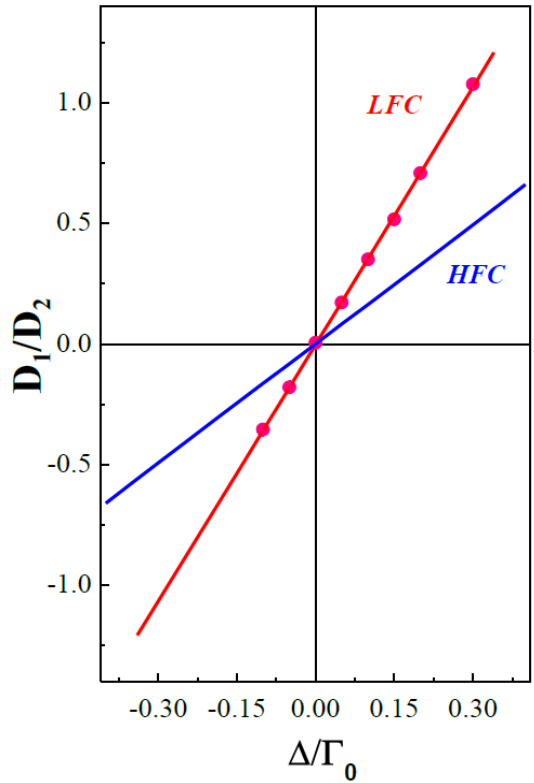}}.\label{fig:3}%
\caption{\textbf{Dependence of the ratio of the amplitudes of the first and
second harmonics of the intensity oscillations on resonant detuning.}
Comparison of the dependence of $D_{1}/D_{2}$ on $\Delta$ for LFC and HFC.
Experimental data for LFC are shown by dots.}%
\end{figure}
\begin{figure}[ptb]
\resizebox{0.4\textwidth}{!}{\includegraphics{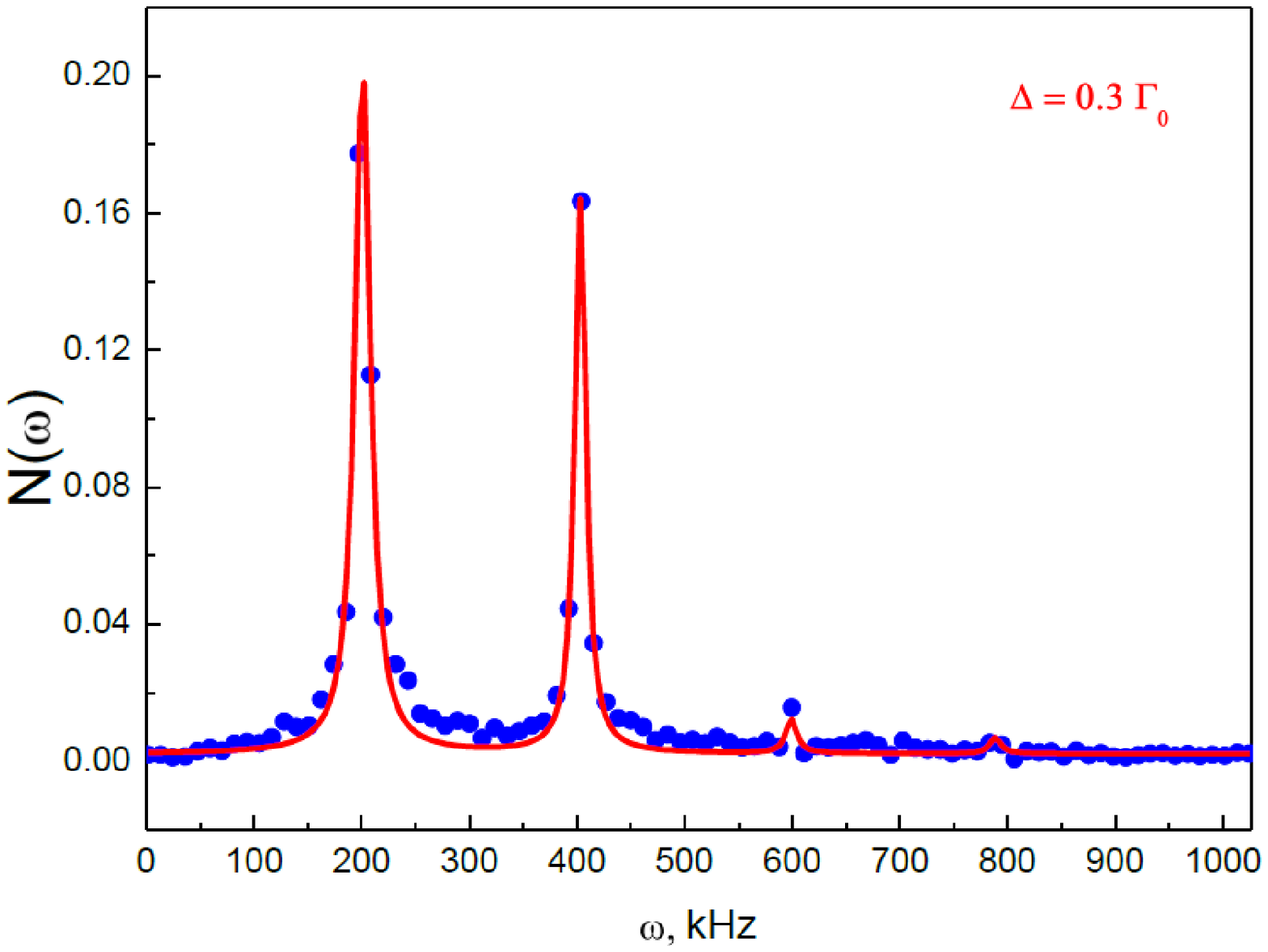}}.\label{fig:4}%
\caption{\textbf{Fourier content of the oscillations.} Fourier content of
the intensity oscillations for $\Delta=0.3\Gamma_{S}$. Other parameters are
defined in the text. Dots correspond to the data, obtained from the Fourier
analysis of the experimentally observed intensity oscillations. Solid line
is the analytical approximation by the set of Lorentzians.}%
\end{figure}

We demonstrated LFC sensitivity in the experiments with the radiation source,
which is a radioactive $^{57}$Co incorporated into rhodium matrix. The source
emits 14.4 keV photons with the spectral width $\Gamma_{S}=1.13$ MHz, which is
mainly defined by the lifetime of 14.4 keV excited state of $^{57}$Fe, the
intermediate state in the cascade decay of $^{57}$Co to the ground state
$^{57}$Fe. The absorber is a 25-$\mu$m-thick stainless-steel foil with a
natural abundance ($\sim$ 2\%) of $^{57}$Fe. Optical depth of the absorber is
$T_{A}=\alpha(0)l=5.18$. The stainless-steel foil is glued on the
polyvinylidene fluoride piezo-transducer that transforms the sinusoidal signal
from radio-frequency generator into the uniform vibration of the foil. The
frequency and amplitude of the sinusoidal voltage were adjusted to have
$\Omega=200$ kHz and $a=5.7$, so that relation $a\Omega\approx\Gamma_{A}$ was
satisfied. The source is attached to the holder of the M\"{o}ssbauer
transducer causing Doppler shift of the radiation field to tune the source in
resonance or out of resonance with the single line absorber. The time
measurements were performed by means of the time--amplitude converter (TAC)
working in the start--stop mode. The start pulses for the converter were
synchronized with radio-frequency generator and the stop pulses were formed
from the signal of 14.4 keV gamma counter at the instant of photon detection
time. A detailed description of the experimental setup is given in
\cite{Vagizov,Shakhmuratov15}.

The experimental results demonstrating the oscillations of the radiation
intensity in time for different values of the resonant detunings $\Delta$ are
shown in Fig. 2. Time dependence of the number of counts is fitted to Eq.
(\ref{Eq2}) (see details in Supplementary information). At exact resonance
($\Delta=0$) only even harmonics are not zero. Time delay of the second
harmonic with respect to the vibration phase is $t_{2}=61$ ns. This delay is
caused by the contribution of dispersion, which produces a phase shift
$2\Omega t_{2}$. The Fourier analysis of the oscillations allows to
reconstruct the dependence of the ratio $D_{1}/D_{2}$ on $\Delta$, which is
shown in Fig. 3. This dependence is compared with that for HFC,
generated by the vibration with high frequency $\Omega=10$ MHz and optimal
value of the modulation index $a=1.08$. We see that LFC is at least two times
more sensitive to resonance than HFC since the slope of the dependence of
$D_{1}/D_{2}$ on $\Delta$ is two times steeper.

Figure 4 shows the Fourier content of the oscillations of the radiation
intensity for LFC when $\Delta=0.3\Gamma_{S}$. The spectrum of these
oscillations contains noticable contributions of the first, second, and third
harmonics. The width of these spectral components is defined by the
length of the time window where the oscillations are measured. In our
experiments the spectral width of each Fourier component is close to 10 kHz.
Thus, we may conclude that within a moderate time of experiment
the proposed method is able to measure the resonant detuning for $^{57}$Fe
with the accuracy of 10 kilohertz, which is 100 times smaller than the
absorption linewidth. This is essentially better accuracy than in the method
of the resonant detuning measurements, used in the gravitational red-shift
measurements \cite{Pound1965,Cranshaw}, which employs four known values of the
calibrated, controllable resonant detunings: two very large, comparable with the
absorption halfwidth, and two very small, comparable but appreciably exceeding
the measured detuning. In time domain measurements, by extending considerably
the length of the time window where the oscillations are collected, one can
reach even higher accuracy of several Hz.

Concluding, we demonstrate a method how with LFC one can measure precisely the
resonant frequency of the absorber with the accuracy equal to a tiny fraction
of the homogeneous absorption width. This method is also applicable in optical
domain. Modulation of the resonant frequency of atoms or impurity ions by
Stark/Zeeman effects or modulation of the frequency of the laser beam by
acousto-optical modulator are equivalent to creation of a frequency comb in a
particular reference frame. The interference of the scattered radiation field
with the incident field is capable to produce the output intensity
oscillations. By a proper choice of the modulation frequency and modulation
index one can make this oscillations to be very sensitive to exact resonance
or to measure the frequency difference between the incident radiation and
resonance frequency of atoms with the accuracy not limited by the value of the
homogeneous linewidth.

\section{METHODS SUMMARY}

The transformation of a spontaneously emitted gamma-radiation on its
propagation through a uniformly vibrating, resonant M\"{o}ssbauer absorber is
described semiclassically in the linear response approximation. Time dependence
of the radiation intensity at the exit of the absorber is studied as a function
of the parameters of the system, such as the frequency and amplitude of modulation,
the detuning of the central frequency of the source from the resonance frequency
of the absorber. We numerically plot the intensity for different parameter values
to demonstrate how the oscillations depend on the resonant detuning. Optimal values
of the modulation index are found for the observation of noticeable effect.

\

\textbf{Acknowledgements} This work was partially funded by the Russian
Foundation for Basic Research (Grant No. 15-02-09039-a), the Program of
Competitive Growth of Kazan Federal University funded by the Russian
Government, the RAS Presidium Programs \textquotedblleft Fundamental optical
spectroscopy and its applications\textquotedblright 
and \textquotedblleft Actual problems of low temperature physics\textquotedblright, 
the National Science Foundation (Grant No. PHY-1307346), and the Robert 
A. Welch Foundation (Award A-1261).

\newpage

\section{Supplementary Information}

In this Supplementary information we give details of mathematical description
of the gamma-radiation intensity oscillations in time for high and low finesse
frequency combs.

\subsection{Introductory remarks}

The propagation of gamma radiation through a resonant M\"{o}ssbauer medium
vibrating with frequency $\Omega$ may be treated classically \cite{Ikonen1985*}
(the references are given at the end of the section in the separate list).
In this approach the radiation field from the source nucleus after passing
through a small diaphragm is approximated as a plane wave propagating along
the direction $\mathbf{x}$. In the coordinate system rigidly bounded to the
absorbing sample, the field, seen by the absorber nuclei, is described by
\begin{equation}
E_{S}(t-t_{0})\propto\theta(t-t_{0})e^{-(i\omega_{S}+\Gamma_{0}/2)(t-t_{0}%
)+ikx+i\varphi(t)}, \label{Eq1*}%
\end{equation}
where $\omega_{S}$ and $k$ are the carrier frequency and the wave number of
the radiation field, $1/\Gamma_{0}$ is the lifetime of the excited state of
the emitting source nucleus, $t_{0}$ is the instant of time when the excited
state is formed, $\Theta(t-t_{0})$ is the Heaviside step-function,
$\varphi(t)=2\pi x_{d}(t)/\lambda=a\sin(\Omega t)$ is a time dependent phase
of the field due to a piston-like periodical displacement of the absorber with
respect to the source, $x_{d}(t)$, and $\lambda$ is the radiation wavelength.

It can be easily shown that radiation intensity at the exit of the vibrating
absorber is the same if the source is vibrated instead of absorber. For
simplicity we consider the vibration of the source with respect to the
absorber and not vice versa, since both cases are equivalent. Then the
radiation field from the source can be expressed as follows%
\begin{equation}
E_{S}(t-t_{0})=E_{C}(t-t_{0})e^{-i\omega_{S}(t-t_{0})+ikx}\sum_{n=-\infty
}^{+\infty}J_{n}(a)e^{in\Omega t}, \label{Eq2*}%
\end{equation}
where $E_{C}(t-t_{0})=E_{0}\theta(t-t_{0})e^{-\Gamma_{0}(t-t_{0})/2}$ is the
common part of the field components, $E_{0}$ is the field amplitude, and
$J_{n}(a)$ is the Bessel function of the $n$th order. The Fourier transform of
this field is%
\begin{equation}
E_{S}(\omega)=E_{0}\sum_{n=-\infty}^{+\infty}\frac{J_{n}(a)e^{in\Omega t_{0}}%
}{\Gamma_{0}/2+i(\omega_{S}-n\Omega-\omega)}, \label{Eq3*}%
\end{equation}
where for shortening of notations the exponential factor with $ikx$ is
omitted. From this expression, it is clear that the vibrating absorber `sees'
the incident radiation as an equidistant frequency comb with spectral
components $\omega_{S}-n\Omega$ whose amplitudes are proportional to
$J_{n}(a)$. Below, for briefness we use the shortened notation $J_{n}%
(a)=J_{n}$.

According to Eq. (\ref{Eq1*}) the intensity of the field%
\begin{equation}
I(t-t_{0})=\left\vert E_{S}(t-t_{0})\right\vert ^{2}=I_{0}\theta
(t-t_{0})e^{-\Gamma_{0}(t-t_{0})},\label{Eq4}%
\end{equation}
where $I_{0}=E_{0}^{2}$, does not oscillate in time. The same result must be
obtained from Eq. (\ref{Eq2*}), which gives%
\begin{equation}
\left\vert E_{S}(t-t_{0})\right\vert ^{2}=\left\vert E_{C}(t-t_{0})\right\vert
^{2}\sum_{n=-\infty}^{+\infty}\sum_{m=-\infty}^{+\infty}J_{n}J_{m}%
e^{i(n-m)\Omega t},\label{Eq5*}%
\end{equation}
where $\left\vert E_{C}(t-t_{0})\right\vert ^{2}=I_{0}(t-t_{0})$ according to
the definition. Therefore, the identity%
\begin{equation}
\sum_{n=-\infty}^{+\infty}\sum_{m=-\infty}^{+\infty}J_{n}J_{m}e^{i(n-m)\Omega
t}=1\label{Eq6*}%
\end{equation}
is to be satisfied. It is consistent with the well known relations between the
Bessel functions (see, for example, \cite{AbramSteg65*}). They are%
\begin{equation}
\sum_{n=-\infty}^{+\infty}J_{n}^{2}=J_{0}^{2}+2\sum_{n=1}^{+\infty}J_{n}%
^{2}=1\label{Eq7*}%
\end{equation}
for zero harmonic ($n=m$), which is the only harmonic giving nonzero
contribution into the radiation intensity in Eq. (\ref{Eq5*}) due to Eq.
(\ref{Eq6*}), and%
\begin{equation}
\sum_{n=-\infty}^{+\infty}J_{n}J_{n+2}=-J_{1}^{2}+2\sum_{n=0}^{+\infty}%
J_{n}J_{n+2}=0,\label{Eq8*}%
\end{equation}
for the second harmonic ($-2\Omega$) with $m=n+2$ in Eq. (\ref{Eq5*}), and%
\begin{equation}
\sum_{n=-\infty}^{+\infty}J_{n}J_{n+4}=J_{2}^{2}-2J_{1}J_{3}+2\sum
_{n=0}^{+\infty}J_{n}J_{n+4}=0,\label{Eq9*}%
\end{equation}
for the fourth harmonic ($-4\Omega$) with $m=n+4$. It can be easily shown that
all even harmonics do not contribute since their amplitudes are zero. As
regards the odd harmonics, they have zero amplitudes because of the
cancelation of the symmetric pairs in their content as, for example, for the
first harmonic ($-\Omega$) with $m=n+1$,%
\begin{equation}
\sum_{n=-\infty}^{+\infty}J_{n}J_{n+1}=(J_{0}J_{1}+J_{-1}J_{0})+(J_{1}%
J_{2}+J_{-2}J_{-1})+...=0,\label{Eq10*}%
\end{equation}
and the third harmonic ($-3\Omega$) with $m=n+3$,%
\begin{equation}
\sum_{n=-\infty}^{+\infty}J_{n}J_{n+3}=(J_{0}J_{3}+J_{-3}J_{0})+(J_{1}%
J_{4}+J_{-4}J_{-1})+...=0.\label{Eq11*}%
\end{equation}
Here the property of the Bessel function, $J_{-n}=(-1)^{n}J_{n}$ (where $n$ is
positive), is taken into account.

If such a field with balanced amplitudes and phases of its harmonics, Eq.
(\ref{Eq2*}), passes through a thick resonant absorber, one may expect that
this balance will be broken and the intensity of the field at the exit of the
absorber will be oscillating.

\subsection{The transformation of the radiation field after passing through a
resonant absorber}

The Fourier transform of the radiation field is changed at the exit of the
resonant absorber as \cite{Monahan*,Vagizov*}
\begin{equation}
E_{out}(\omega)=E_{0}\sum_{n=-\infty}^{+\infty}\frac{J_{n}\exp\left[  in\Omega
t_{0}-\frac{b}{\Gamma_{A}/2+i(\omega_{A}-\omega)}\right]  }{\Gamma
_{0}/2+i(\omega_{S}-n\Omega-\omega)}, \label{Eq12*}%
\end{equation}
where $\omega_{A}$ and $\Gamma_{A}$ are the frequency and linewidth of the
absorber, $b=T_{A}\Gamma_{0}/4$ is the parameter depending on the effective
thickness of the absorber $T_{A}=fn\sigma$, $f$ is the Debye-Waller factor,
$n$ is the number of $^{57}$Fe nuclei per unit area of the absorber, and
$\sigma$ is the resonance absorption cross section. The source linewidth
$\Gamma_{S}$ can be different from $\Gamma_{0}$ due to the contribution of the
environment of the emitting nucleus in the source. In this case $\Gamma_{0}$
can be simply substituted by $\Gamma_{S}$ in Eq. (\ref{Eq12*}).

Time dependence of the amplitude of the output radiation field is found by
inverse Fourier transformation%
\begin{equation}
E_{out}(t-t_{0})=\frac{1}{2\pi}\int_{-\infty}^{+\infty}E_{out}(\omega
)e^{-i\omega(t-t_{0})}d\omega. \label{Eq13*}%
\end{equation}
Then, the intensity of the field is%
\begin{equation}
I_{out}(t-t_{0})=\frac{1}{(2\pi)^{2}}\int_{-\infty}^{+\infty}d\omega_{1}%
\int_{-\infty}^{+\infty}d\omega_{2}E_{out}(\omega_{1})E_{out}^{\ast}%
(\omega_{2})e^{i(\omega_{2}-\omega_{1})(t-t_{0})}. \label{Eq14*}%
\end{equation}

In time domain experiments the phase of the vibrations is fixed but emission
time of gamma-photons is random. Therefore, the observed radiation intensity
is averaged over $t_{0}$
\begin{equation}
\left\langle I_{out}(t-t_{0})\right\rangle _{t_{0}}=\lim_{T\rightarrow\infty
}\int_{-T}^{t}I_{out}(t-t_{0})dt_{0}. \label{Eq15*}%
\end{equation}
Calculation of this integral gives for the number of gamma-photon counts at
the exit of the absorber, $N_{out}(t)\propto\left\langle I_{out}%
(t-t_{0})\right\rangle _{t_{0}}$, the following expression \cite{Vagizov*}%
\begin{equation}
N_{out}(t)/N_{0}=\sum_{n,m=-\infty}^{+\infty}J_{n}J_{m}e^{i(n-m)\Omega
t}B_{nm}(\Delta), \label{Eq16*}%
\end{equation}
where $N_{0}$ is the number of counts far from resonance and
\begin{equation}
B_{nm}(\Delta)=\frac{\Gamma_{S}}{2\pi}\int_{-\infty}^{+\infty}\frac
{e^{-\frac{b}{\Gamma_{A}/2+i(\Delta+n\Omega-\omega)}-\frac{b}{\Gamma
_{A}/2-i(\Delta+m\Omega-\omega)}}}{(\Gamma_{S}/2)^{2}+\omega^{2}}d\omega,
\label{Eq17*}%
\end{equation}
where $\Delta=\omega_{A}-\omega_{S}$ is the resonance detuning of the source
and absorber. In derivation of Eq. (\ref{Eq17*}) the substitution
$\omega^{\prime}=\omega-\omega_{S}+n\Omega$ is used in Eqs. (\ref{Eq12*}) and
(\ref{Eq14*}). Then the prime is omitted.

\subsection{Intensity oscillations}

To analyze the oscillations of the radiation intensity after passing through
the vibrated absorber it is convenient to group the terms in Eq. (\ref{Eq16*})
as follows%
\begin{equation}
N_{out}(t)/N_{0}=\mathtt{I}_{0}(\Delta)+2\operatorname{Re}\sum_{n=1}^{+\infty
}\mathtt{I}_{n}(\Delta)e^{-in\Omega t}, \label{Eq18*}%
\end{equation}
where $\mathtt{I}_{n}(\Delta)$ is the $n$th-harmonic amplitude of the
radiation intensity oscillations at the exit of the absorber. The amplitudes
of the harmonics are defined by the products of the radiation amplitudes of
the frequency comb (\ref{Eq12*}), transformed by the absorber. For example, the
amplitude of zero harmonic is%
\begin{equation}
\mathtt{I}_{0}(\Delta)=J_{0}^{2}B_{00}(\Delta)+\sum_{n=1}^{+\infty}J_{n}%
^{2}\left[  B_{nn}(\Delta)+B_{(-n)(-n)}(\Delta)\right]  , \label{Eq19*}%
\end{equation}
where the coefficients $B_{00}(\Delta)$, $B_{nn}(\Delta)$, and $B_{(-n)(-n)}%
(\Delta)$ are transmitted intensities of $0$, $n$, and $-n$ components of the
incident comb (\ref{Eq2*}). They are
\begin{equation}
B_{00}(\Delta)=\frac{\Gamma_{S}}{2\pi}\int_{-\infty}^{+\infty}\frac
{e^{-\frac{b\Gamma_{A}}{(\Gamma_{A}/2)^{2}+(\Delta-\omega)^{2}}}}{(\Gamma
_{S}/2)^{2}+\omega^{2}}d\omega, \label{Eq20*}%
\end{equation}%
\begin{equation}
B_{(\pm n)(\pm n)}(\Delta)=\frac{\Gamma_{S}}{2\pi}\int_{-\infty}^{+\infty
}\frac{e^{-\frac{b\Gamma_{A}}{(\Gamma_{A}/2)^{2}+(\Delta\pm n\Omega
-\omega)^{2}}}}{(\Gamma_{S}/2)^{2}+\omega^{2}}d\omega. \label{Eq21*}%
\end{equation}
Thus, $\mathtt{I}_{0}(\Delta)$ is just the sum of the transmitted intensities
of all spectral components of the frequency comb (\ref{Eq2*}).

The first harmonic%
\begin{equation}
\mathtt{I}_{1}(\Delta)=\sum_{n=0}^{+\infty}J_{n}J_{n+1}\left[  B_{n(n+1)}%
(\Delta)-B_{(-n-1)(-n)}(\Delta)\right]  , \label{Eq22*}%
\end{equation}
contains the difference of two terms originating from the interference of
two neighboring components of the frequency comb $\pm n$ and $\pm(n+1)$. They
are red (for sign $+$) and blue (for sign $-$) detuned from resonance. This
difference is%
\begin{multline}
B_{n(n+1)}(\Delta)-B_{(-n-1)(-n)}(\Delta)=\\
\frac{\Gamma_{S}}{2\pi}\int_{-\infty}^{+\infty}\frac{e^{-\frac{b}{\Gamma
_{A}/2+i(\Delta+n\Omega-\omega)}-\frac{b}{\Gamma_{A}/2-i[\Delta+(n+1)\Omega
-\omega]}}-e^{-\frac{b}{\Gamma_{A}/2+i[\Delta-(n+1)\Omega-\omega]}-\frac
{b}{\Gamma_{A}/2-i(\Delta-n\Omega-\omega)}}}{(\Gamma_{S}/2)^{2}+\omega^{2}%
}d\omega. \label{Eq23*}%
\end{multline}
It is easy to show (by substitution $\omega=-\omega^{\prime}$ in the second
exponent) that the difference of the interference terms is zero if $\Delta=0$.

The second and third harmonics are described by equations%
\begin{equation}
\mathtt{I}_{2}(\Delta)=-J_{1}^{2}B_{-11}(\Delta)+\sum_{n=0}^{+\infty}%
J_{n}J_{n+2}\left[  B_{n(n+2)}(\Delta)+B_{(-n-2)(-n)}(\Delta)\right]  ,
\label{Eq24*}%
\end{equation}%
\begin{equation}
\mathtt{I}_{3}(\Delta)=-J_{1}J_{2}[B_{-12}(\Delta)-B_{-21}(\Delta)]+\sum
_{n=0}^{+\infty}J_{n}J_{n+3}\left[  B_{n(n+3)}(\Delta)-B_{(-n-3)(-n)}%
(\Delta)\right]  , \label{Eq25*}%
\end{equation}
which contain the interference terms of the frequency-comb amplitudes with
$m-n=2$ for $\mathtt{I}_{2}(\Delta)$ and $m-n=3$ for $\mathtt{I}_{3}(\Delta)$
[see Eq. (\ref{Eq17*})]. The third harmonic is zero if $\Delta=0$ because it
contains the difference of the interference terms, while the second harmonic
is not zero since it is the sum of the interference terms.

The coefficients $D_{n}$ in equation (2) of the main part of the manuscript,
describing the number of count oscillations in time, are related to the
harmonics $\mathtt{I}_{n}(\Delta)$ as $D_{0}=\mathtt{I}_{0}(\Delta)$,
$D_{2n}=2\left\vert \mathtt{I}_{2n}(\Delta)\right\vert $, and $D_{2n+1}%
=2S(\Delta)\left\vert \mathtt{I}_{2n+1}(\Delta)\right\vert $, where
$S(\Delta)=\Delta/\left\vert \Delta\right\vert $. The phases of the harmonics
are defined as $\Omega t_{n}=\arctan\left[  \operatorname{Im}\mathtt{I}%
_{n}(\Delta)/\operatorname{Re}\mathtt{I}_{n}(\Delta)\right]  $.

\subsection{High finesse frequency comb}

If the modulation frequency is much larger than the absorption linewidth
($\Omega\gg\Gamma_{A}$) and the central frequency of the comb $\omega_{S}$ is
close to the resonant frequency of the absorber $\omega_{A}$ ($\left\vert
\Delta\right\vert <\Gamma_{A}$), then only the central frequency of the comb
is changed after passing through a thick absorber. Therefore, one may expect
that in Eq. (\ref{Eq16*}) only the components $B_{00}$, $B_{0n}$ and $B_{n0}$
become different from $1$, while others are almost unity since for them the
exponents in the integral (\ref{Eq17*}) are unity if the condition $n\Omega\gg
b$ is satisfied. In this case we can use approximate equations%
\begin{equation}
\mathtt{I}_{0}(\Delta)\approx1-J_{0}^{2}[1-B_{00}(\Delta)], \label{Eq26*}%
\end{equation}%
\begin{equation}
\mathtt{I}_{1}(\Delta)\approx J_{0}J_{1}\left[  B_{01}(\Delta)-B_{-10}%
(\Delta)\right]  , \label{Eq27*}%
\end{equation}%
\begin{equation}
\mathtt{I}_{2}(\Delta)\approx J_{0}J_{2}\left[  B_{02}(\Delta)+B_{-20}%
(\Delta)-2\right]  , \label{Eq28*}%
\end{equation}
which are derived taking into account the relations (\ref{Eq7*}), (\ref{Eq8*}),
and (\ref{Eq10*}). The product $J_{0}J_{1}=J_{0}(a)J_{1}(a)$ in Eq. (\ref{Eq27*})
has global maximum when the modulation index is $a=1.08$. Therefore, for
nonresonant excitation ($\Delta\neq0$ and $\left\vert \Delta\right\vert
<\Gamma_{A}$) the first harmonic of the intensity oscillations has maximum
amplitude for this value of the modulation index. Since it is not large, we
may approximate the intensity oscillations taking into account only three
harmonics $n=0,1,2$ in Eq. (\ref{Eq18*}).
\begin{figure}[ptb]
\resizebox{0.5\textwidth}{!}{\includegraphics{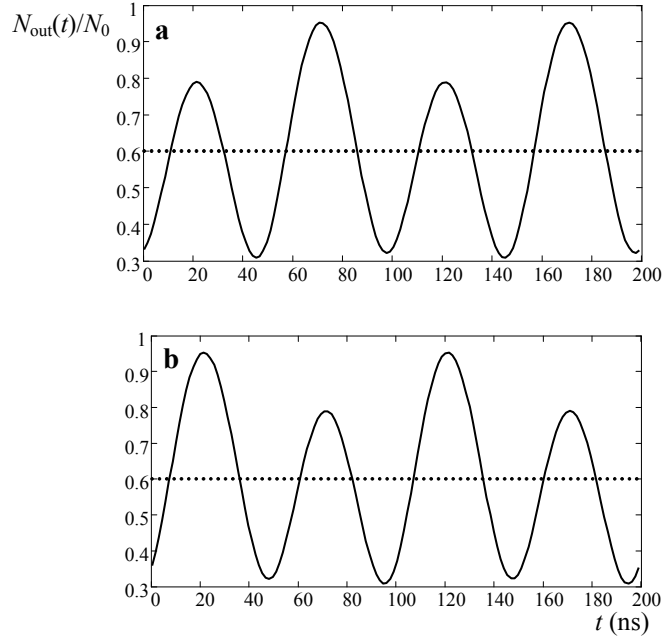}}\caption{Oscillations
of the radiation intensity for high finesse comb ($\Omega$=10 MHz) are shown by
solid lines. The resonant detuning is $\Delta=-200$ kHz in (\textbf{a}) and $\Delta=200$ kHz in (\textbf{b}). Dotted line shows the level of $\mathtt{I}_{0}(\Delta)$. The value of the parameters are given in the text.}%
\label{fig:5}%
\end{figure}

However, to achieve high accuracy we have to take into account also the
contribution of two spectral components of the comb, neighboring the resonant
component \cite{Shakhmuratov15*}. This is because far wings of the Lorentzian
line give small, but noticable contribution. In our case, when the central
component of the comb is in resonance, these nearest components are $+\Omega$
and $-\Omega$. Then, for example, $\mathtt{I}_{0}(\Delta)$ is modified due to
this small nonresonant contribution as%
\begin{equation}
\mathtt{I}_{0}(\Delta)=1-J_{0}^{2}[1-B_{00}(\Delta)]-J_{1}^{2}[2-B_{11}%
(\Delta)-B_{-1-1}(\Delta)]. \label{Eq29*}%
\end{equation}
The value of the correction due to the additional terms is about 2\% if
$\Omega/\Gamma_{0}=10$, $\Delta=0.2\Gamma_{0}$, and $T_{A}=5$.

In conclusion of this section, we show in Fig. 5 two examples of the intensity
oscillations (not approximated) for $\Gamma_{A}=\Gamma_{S}=\Gamma_{0}=1.13$
MHz, $T_{A}=5.2$, $\Omega=10$ MHz, $a=1.08$, and $\Delta=\pm200$ kHz. The
dependence of $D_{1}/D_{2}$ on $\Delta$ (not approximated) is shown in Fig. 3
of the main part of the manuscript.

\newpage
\subsection{Low finesse comb}

For the low finesse comb the modulation frequency is much smaller than the
absorption linewidth ($\Omega\ll\Gamma_{A}$). If the central frequency of the
comb $\omega_{S}$ is close to the resonant frequency of the absorber
$\omega_{A}$ ($\left\vert \Delta\right\vert <\Gamma_{A}$), then many spectral
components of the comb are changed after passing through a thick absorber.
Therefore, to describe the oscillation of the output radiation intensity we
have to take many terms in the equations (\ref{Eq19*}), (\ref{Eq22*}),
(\ref{Eq24*}), and (\ref{Eq25*}) for $\mathtt{I}_{n}(\Delta)$, $n=0,1,2,3$.

Intuitively, one may expect that LFC sensitivity is maximal if all noticeable
components of the comb are modified after passing through the absorber. This
takes place if the product $a\Omega$, which specifies the total spectral width
of the comb or the frequency range, covered by the comb components with noticeable
amplitudes, is close to the width of the absorption line. Mathematically, this
expectation can be verified by plotting the amplitude $D_{1}=2S(\Delta
)\left\vert \mathtt{I}_{1}(\Delta)\right\vert $ versus modulation index $a$
for a fixed values of the modulation frequency $\Omega$ and resonant detuning
$\Delta$. Figure 6 shows the dependence of $D_{1}$ on the modulation index $a$
for $\Omega=\Delta=200$ kHz and $T_{A}=5.2$. The amplitude $D_{1}$ takes
maximum value when $a=7$. For this value the product $a\Omega$ is close to the
linewidth of the absorber $\Gamma_{A}$.
\begin{figure}[ptb]
\resizebox{0.5\textwidth}{!}{\includegraphics{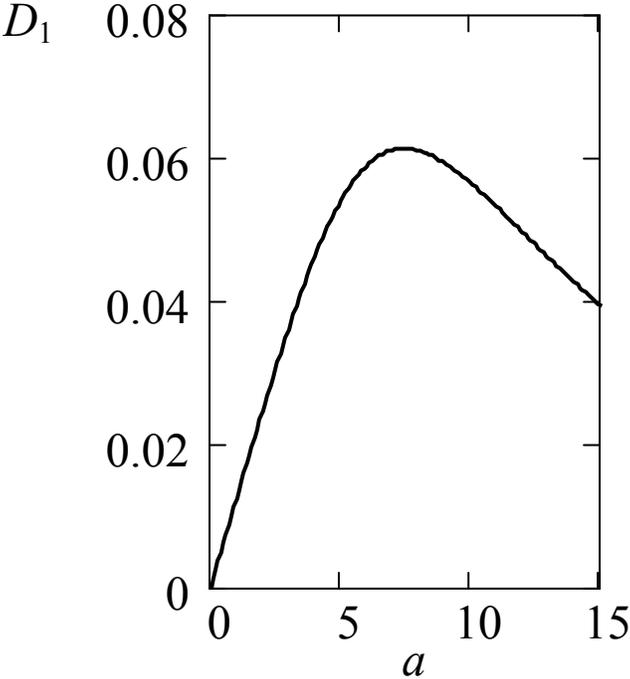}}\caption{The
dependence of the first-harmonic amplitude $D_{1}$ of the intensity
oscillations on the modulation index $a$. The modulation frequency and
resonant detuning are $\Omega=\Delta=200$ kHz.}%
\label{fig:6}%
\end{figure}

\end{document}